\documentclass[prx,aps,twocolumn]{revtex4-1}
\usepackage{amsmath,amssymb, graphicx}
\usepackage{bm,times, color}
\usepackage[colorlinks]{hyperref}
\usepackage[svgnames]{xcolor}
\hypersetup{linkcolor=DarkBlue, citecolor=DarkBlue, filecolor=black, urlcolor=DarkBlue}
\usepackage{comment, mathrsfs,slashed}
\def \p{\partial}

\def \mb{\mathbf}

\def \lan{\langle}
\def \ran{\rangle}

\begin{document}
\preprint{EFI-xx}
\title{Self-dual $\nu=1$ bosonic quantum Hall state in mixed-dimensional QED}

\author{Wei-Han Hsiao and Dam Thanh Son}
\affiliation{Kadanoff Center for Theoretical Physics, University of Chicago,
Chicago, Illinois 60637, USA}
\date{September 2018}

\begin{abstract}
We consider a (2+1)-dimensional Wilson-Fisher boson coupled to a (3+1)-dimensional U(1) gauge field. This theory possesses a strong-weak
duality in terms of the coupling constant $e$ and is self-dual at a
particular value of $e$. We derive exact relations
between transport coefficients for a $\nu=1$ quantum Hall state at the
self-dual point. Using boson-fermion duality, we map the $\nu = 1$
bosonic quantum Hall state to a Fermi sea of the dual fermion and
observe that the exact relationships between transport coefficients
at the bosonic self-dual point are reproduced by a simple random-phase
approximation (RPA), coupled with a Drude formula, in the fermionic theory.
We explain this success of the RPA by pointing out a cancellation of a
parity-breaking term in the fermion theory which occurs only at the
self-dual point, resulting in the fermion self-dual theory explored
previously. In addition, we argue that the equivalence of the self-dual
structure can be understood in terms of electromagnetic duality or
modular invariance, and these features are not inherited by the
non-relativistic cousins of the present model.
\end{abstract}

\maketitle

\section{Introduction}

Dualities provide powerful tools for the study of phases of matter
and emergent properties of strongly correlated systems, often allowing
one to derive result not accessible by perturbative methods.
Particularly useful are strong-weak dualities and self-dualies, which,
in many cases, provide quantitative predictions in addition to
qualitative insights. Well known examples include the two-dimensional
classical Ising model and the one-dimensional quantum Ising
model~\cite{RevModPhys.51.659}. In both systems, the location of the
phase trasition can be found from the condition of self-duality.

Recently, field-theoretical infrared dualities in 2+1 dimensions have
attracted the attention of the condensed-matter and high-energy
communities. These dualities are the relativistic version of
non-relativistic flux attachment and can also be regarded as a
higher-dimensional generalization of (1+1)-dimensional bosonization
\cite{PhysRevD.11.2088, PhysRevD.11.3026}.  A large number of
equivalent pairs of field theories can be derived from a single
``seed'' boson-fermion
duality~\cite{SEIBERG2016395,PhysRevX.6.031043}.  Models in this web
of dualities have found applications in the $\nu=\frac12$ quantum Hall
problem \cite{PhysRevX.5.031027}, the strongly interacting surface
states of topological insulators
\cite{PhysRevX.5.041031,PhysRevB.93.245151}, and the deconfined
quantum critical points \cite{PhysRevX.7.031051,PhysRevX.7.031052}.
Progress has been made in trying to come up with a microscopic derivation of the seed duality \cite{PhysRevD.94.085009,PhysRevLett.118.011602,PhysRevLett.120.016602} and in extracting quantitative predictions from the dualities \cite{PhysRevB.94.214415,PhysRevB.96.075127}. In particular, in our previous work \cite{PhysRevB.96.075127}, we made use of the self-duality property of a certain theory to derive constraints on the physical response at the self-dual point.

The goal of this paper is to extend the method of
Ref.~\cite{PhysRevB.96.075127} to bosonic quantum Hall states.  Our 
model consists of a (2+1)-dimensional [(2+1)D] Wilson-Fisher boson coupled to a U(1) gauge field propagating in (3+1) dimensions, and is self dual at a certain value of the coupling constant. The self duality allows one, in analogy with the fermionic case considered in Ref.~\cite{PhysRevB.96.075127}, to derive nontrivial relations between transport coefficients at quantum Hall filling factor $\nu=1$.  These nontrivial relations, Eqs.~(\ref{result4}), (\ref{result5}), and (\ref{result6}) below, include a semicircle law for the conductivities, a relationship between the thermal Hall angle and the Hall angle, and a relationship between the bulk thermal Hall conductivity and the thermoelectric coefficients.

The $\nu=1$ bosonic quantum Hall state allows an interpretation in terms of a Fermi liquid~\footnote{Strictly speaking, a marginal Fermi liquid.} of composite fermions.  Therefore, one may pose the question. What are the properties of the composite fermions that guarantee the self-dual properties of the bosonic theory?  In this paper we will show that it is a discrete symmetry, which we call $\mathsf{T}_{\!\rm f}$, which acts on the composite fermion like a time reversal. The simplest random-phase approximation (RPA) preserves $\mathsf{T}_{\!\rm f}$ at any value of the gauge coupling~\cite{PhysRevB.95.045118}, but beyond RPA the symmetry is realized only at the self-dual value of the coupling. 

This paper is structured as follows. In Sec. \ref{Building Blocks},
we first introduce our notations for the building blocks that will be
used later for constructing the action.  In Sec. \ref{selfdualQED}, we review the self-dualities studied in
Ref.~\onlinecite{PhysRevB.96.075127} in a manner that incorporates
both bosonic and fermionic particle-vortex dualities. In
Sec. \ref{BosonicHallState}, we discuss how fractional bosonic
quantum Hall states and the self-dual structures can be understood in
terms of the composite fermions. In particular, we focus on the $\nu
=1$ state and examine how transport properties at the self-dual point
can be understood using a simple fermionic picture. It is found that
there the self-dual point on the bosonic side corresponds to a self-dual,
time-reversal symmetric point on the fermionic side, and we
again elaborate its relation with electromagnetic duality. Such
a simultaneously self-dual and time-reversal-symmetric structure can be
understood in terms of modular coupling constant and its
transformation under PSL(2,$\mathbb Z$). At the end, we give a short
summary and discuss some open directions.

\section{Building Blocks}\label{Building Blocks}

In the field theories that we will consider, the matter field will be
a (2+1)-dimensional $O(2)$ Wilson-Fisher (WF) boson, denoted by $\phi$
or $\tilde{\phi}$, or a two-component Dirac fermion denoted by $\psi$
or $\chi$. When minimally coupled to gauge fields $A$, their actions
are abbreviated as follows:
\begin{align}
\label{WF}& I_{\rm m}[\phi, A] = |D_A\phi|^2-|\phi|^4,\\
\label{Dirac}& I_{\rm m}[\psi, A] = i\bar{\psi}\slashed{D}_A\psi,
\end{align}
where $D_A=\partial-iA$. To write formulas that can be applied equally
to the bosonic and fermionic cases, we will denote the matter field by
$\Psi_p$, where $p=1$ corresponds to the WF boson and $p=2$
corresponds to the Dirac fermion: $\Psi_1=\phi$, $\Psi_2=\psi$.

For gauge fields, in this work we mainly consider three kinds of
actions: a (3+1)D Maxwell term $I_{\rm M}$ and two types of (2+1)D Chern-Simons (CS) terms, $I_{\rm BF}$ and $I_{\rm CS}$. They are given as follows:
\begin{align}
  \label{Maxwellterm} I_{\rm M}[A; e] &= -\frac1{4e^2} \!\int\! d^4x\, F_{\mu\nu}F^{\mu\nu},\\
\label{BFterm} I_{\rm BF}[a,A] &= \phantom{+}\frac1{2\pi}\!\int\! d^3x\, \epsilon^{\mu\nu\lambda}a_{\mu}\p_{\nu}A_{\lambda},\\
\label{CSterm} I_{\rm CS}[A] &= \phantom{+} \frac1{4\pi}\!\int\! d^3x\, \epsilon^{\mu\nu\lambda}A_{\mu}\p_{\nu}A_{\lambda}.
\end{align}
Note that the Maxwell term involves integration over a (3+1)-dimensional
space-time.  On some occasions, it is convenient to consider its
reduction to a nonlocal term in 2+1 dimensions \cite{Marino:1992xi}:
\begin{align}
  \label{PQEDterm}& I_{\rm P}[A;e] =
  -\frac i{2e^2}\!\int\! d^3x\, d^3x'\,
  F_{\mu\nu}(x)\frac1{\sqrt{-\p^2}}F^{\mu\nu}(x').
\end{align}
Finally, the action $I_{\theta}[A, \Delta\theta]$ refers to an axion
action with a $\theta$-angle difference $\Delta\theta$ across a given
domain wall:
\begin{equation}
  I_\theta[A, \Delta\theta]  = \frac{1}{32\pi^2}\int d^4x\, \theta(z)\epsilon^{\mu\nu\lambda\rho}F_{\mu\nu}F_{\lambda\rho},
\end{equation}
where $\theta(z)$ has a jump at $z=0$, $\theta(+\epsilon)-\theta(-\epsilon)=\Delta\theta$, and is constant elsewhere.

\section{Review of Self-Dual Mix-Dimensional QED}\label{selfdualQED}

\subsection{Dualities}

In this section we review self-dual theories involving a (2+1)-dimensional field theory on a $z=0$ brane coupled to a gauge field
\eqref{Maxwellterm} propagating in (3+1)-dimensional Minkowski
space-time. First, let us recall the
bosonic~\cite{PESKIN1978122,PhysRevLett.47.1556, PhysRevB.39.2756} and
fermionic particle-vortex
dualities~\cite{PhysRevX.5.031027,PhysRevX.5.041031,PhysRevB.93.245151}.
In our notation, these dualities can be written in a uniform way,
\begin{equation}\label{pvduality}
  I_{\rm m}[\Psi_p, A] \leftrightarrow I_{\rm m}[\tilde{\Psi}_p, a]+\frac{1}{p}I_{\rm BF}[a,A].
\end{equation}

In Eq.~\eqref{pvduality}, $A_{\mu}$ is considered a background (probe)
field. We add to both sides
of Eq.~\eqref{pvduality} a (3+1)D Maxwell term~\eqref{Maxwellterm} and
integrate over $A_{\mu}$.  The duality now becomes
\begin{equation}\label{pvduality-IM}
  I_{\rm m}[\Psi_p, A] + I_{\rm M}[A;e]\leftrightarrow I_{\rm m}[\tilde{\Psi}_p, a]+\frac{1}{p}I_{\rm BF}[a,A] + I_{\rm M}[A;e]
\end{equation}

This procedure introduces an exactly marginal coupling $e$ into the
theory.  The theories on the two sides of Eq.~(\ref{pvduality}) are now
of the ``brane-world'' type, with matter living on a (2+1)D ``brane''
interacting with a (3+1)D ``bulk'' field.  This type of theory can be
regarded as a Lorentz-invariant version of condensed-matter systems,
such as graphene or a superconducting thin film, with noninstantaneous Coulomb
interaction.  In the rest of the paper, we often call the tilde
variables $\tilde{\Psi}_p$ the ``composite particles'' or ``vortices''
of the ordinary matter fields $\Psi_p$.

The operator mapping between the two sides of the duality can be found
by equating the variations of the two sides of Eq.~\eqref{pvduality}
with respect to $A$,
and from the equation of motion obtained by varying the action on the right
hand side of Eq.~(\ref{pvduality}) with respect to $a$,
\begin{align}
  \label{dualitymapping1} J^{\mu}_p &= \phantom{+}\frac{1}{2\pi p}\epsilon^{\mu\nu\lambda}\p_{\nu}a_{\lambda},\\
  \label{dualitymapping2} \tilde{J}^{\mu}_p &= -\frac{1}{2\pi p}\epsilon^{\mu\nu\lambda}\p_{\nu}A_{\lambda}.
\end{align}
The $\mu = 0$ component reads \footnote{We use the $(+--)$ signature for
  the metric and the convention of Ref.~\cite{LL2}; in particular,
  $F_{12} = -B$ and $F_{0i}=E^i$.}
\begin{align}\label{dualitymappingrho}
  \rho_p = -\frac b{2\pi p},\quad \tilde\rho_p = \frac B{2\pi p}\,.
\end{align}
The implication of Eq.~(\ref{dualitymappingrho}) is that a particle on
one side of the duality maps to a flux on the other side.  Moreover,
the flux is double in the fermionic case ($p=2$).  Next we can also
look at the relationship between the filling fraction $\nu =
2\pi\rho/B$ on the two sides.  From Eq.~\eqref{dualitymappingrho} we
find
\begin{align}\label{fillingmapping}
  \nu \tilde{\nu} = \frac{2\pi \rho_p}B\,\frac{2\pi\tilde\rho_p}b =
  -\frac1{p^2}.
\end{align}
Similarly, the spatial components of Eqs.~\eqref{dualitymapping1}
and~\eqref{dualitymapping2} read
\begin{align}\label{dualitymappingj}
  J^i_p = -\frac1{2\pi p}\epsilon^{ij}e^j,\ \
  \tilde J^i_p = \frac1{2\pi p}\epsilon^{ij}E^j,
\end{align}
where $\epsilon^{ij}$ is the antisymmetric tensor, $\epsilon^{12} =
-\epsilon^{21} =1$.

One can take the Gaussian integral over $A$ on the right-hand side (or
the vortex side) of \eqref{pvduality}.  The result is
\begin{equation}
  \label{selfpvdual} I_{\rm m}[\Psi_p, A]+I_{\rm M}[A;e]\leftrightarrow
  I_{\rm m}[\tilde{\Psi}_p, a]+I_{\rm M}[a, \tilde e],
\end{equation}
where 
\begin{equation}\label{selfpvdual-e}
   \tilde e = \frac{4\pi p}{e}
\end{equation}

The duality is now a strong-weak duality and is reminiscent of
electromagnetic duality. In particular, there exists a self-dual point
$e^2 = 4\pi p$ at which both sides of the theories become exactly
the same. The latter should be of no surprise since the coupling $a\,
d A$ is a $\mathsf S$ operation on a three-dimensional conformal field theory (CFT)
\cite{1126-6708-1999-04-021,Witten:2003ya}. It also helps explain the
existence of the strong-weak duality as a legacy of the modular
symmetry.

One can use duality to constrain the physics at finite chemical
potential and magnetic field.
According to Eq.~\eqref{fillingmapping}, electromagnetic duality maps
a state with filling factor $\nu$ to a state with filling factor
$\tilde{\nu}=-(p^2\nu)^{-1}$.  In particular, in the fermionic case
$p=2$, dualty maps $\nu=1/2$ to $\tilde\nu=-1/2$; that is, it maps a
filled zeroth Landau level to an empty one \footnote{ The spectrum of the Dirac fermion in a magnetic field contains a Landau level at zero energy. The $\nu = 1/2$ state here refers to the state filling this 0th Landau level. The non-relativistic filling factor $\nu_{\rm NR} = \frac{1}{2}$ corresponds to filling half of the 0th Landau level $\nu = 0$ in this context.}.  Combined with time reversal,
the duality mapping can relate the physics at filling factors $\nu$ and
$-\tilde\nu=(p^2\nu)^{-1}$.  In particular, at $\nu=\pm p^{-1}$ and
when the coupling constant is tuned to the self-dual value, duality
combined with time reversal maps the theory to itself, a fact that we
will explored in the next section.

\subsection{Transport coefficients}\label{selfdualtransport}

We now use duality to put constraints on transport properties.  They
include the electronic conductivity $\sigma$, thermoelectric tensor
$\alpha$, thermal conductivity in the absence of electric field
$\bar{\kappa}$, and thermal conductivity in the absence of electric
current $\kappa$. They are defined by the following identities (with
$p$ subscript suppressed):
\begin{subequations}
\begin{align}
& J^i = \sigma^{ij}E^j +\alpha^{ij}\p_jT,\\
\label{heatcurrent} & q^i = -T\alpha^{ij}E^j-\bar{\kappa}^{ij}\p_jT,\\
& \kappa = \bar{\kappa}-T\alpha\sigma^{-1}\alpha,
\end{align}
\end{subequations}
where $q^i$ is the heat current.  Similar equations apply to the dual
theory, where all quantities are replaced by their version with a tilde,
\begin{subequations}
\begin{align}
& \tilde J^i = \tilde \sigma^{ij} \tilde E^j + \tilde \alpha^{ij}\p_jT\\
\label{heatcurrent} & q^i = -T \tilde \alpha^{ij} \tilde E^j - \tilde{\bar{\kappa}}^{ij}\p_jT\\
& \tilde \kappa = \tilde{\bar{\kappa}}-T\tilde \alpha \tilde \sigma^{-1} \tilde \alpha.
\end{align}
\end{subequations}
Note that the temperature $T$ and the heat current $q^i$ are invariant
under electromagnetic duality.  First, let us look at the $J$
equation. Using the duality mappings \eqref{dualitymappingj}, it can
be shown that
\begin{subequations}\label{result123}
\begin{align}
\label{result1} & \sigma = -\frac1{(2\pi p)^2}\epsilon\tilde{\sigma}^{-1}\epsilon\\
\label{result2}& \alpha = (2\pi p)\sigma\epsilon\tilde\alpha = \frac{1}{(2\pi p)}\epsilon\tilde{\sigma}^{-1}\tilde{\alpha}\\
\label{result3} & \bar{\kappa} = \tilde{\bar{\kappa}}-T\tilde{\alpha}\tilde{\sigma}^{-1}\tilde{\alpha} = \tilde{\kappa}.
\end{align}
\end{subequations}
Identities similar to Eqs.~\eqref{result123} are explicitly found in
Ref.~\cite{Donos2017}, where hydrodynamics equipped with bulk
electromagnetic duality is studied. In the context of holographic duality, similar results, but with
different numerical factors, are found in
Ref.~\cite{PhysRevD.76.106012}.

Equations \eqref{result123} become much more restrictive when a
self-dual structure is present.  Let us now consider the self-dual
value $e^2=4\pi p$ and self-dual configuration $\nu = \pm p^{-1}$, and
for simplicity we consider the response at infinite wavelength and finite
frequency. The conductivity tensor $\sigma$, due to rotational
invariance, must have the form
 \begin{equation}
   \sigma^{ij} = \sigma_{xx}\delta^{ij} + \sigma_{xy} \epsilon^{ij}
\end{equation}
Under duality, since $\nu$ flips sign, $\sigma$ changes to its
transpose. For instance, $\tilde{\sigma} = \sigma^T$. Consequently, we
can further simplify Eqs.~(\ref{result123}) and conclude that \
\begin{align}
\label{result4}& \sigma_{xx}^2+\sigma_{xy}^2 = \frac{1}{(2\pi p)^2}\\
\label{result5}& \frac{\alpha_{xy}}{\alpha_{xx}} = \tan\Big(\frac{\pi}{4}+\frac{\theta_H}{2}\Big),\quad \theta_H = \mathrm{arctan}\frac{\sigma_{xy}}{\sigma_{xx}}\\
\label{result6} & \bar{\kappa}_{xy} = -\kappa_{xy} = \frac{T}{2}\frac{\alpha_{xx}^2+\alpha_{xy}^2}{1/(2\pi p)}.
\end{align}
The Hall angle $\theta_H$ is a function of $\omega/T$, where $T$ is the temperature. Although we cannot compute $\theta_H$ at arbitrary $T$, there are two limit cases to be illustrated. In the clean ballistic limit $\omega/T\to\infty$, $\sigma_{xx} = 0$ and $\sigma_{xy} = \frac{1}{2\pi p}\Rightarrow \theta_H = \frac{\pi}{2}.$ In the opposite hydrodynamic limit $T/\omega\to \infty$, $\sigma_{xy} = 0$, and thus $\theta_H = 0$.
These are the main results obtained in the previous work
\cite{PhysRevB.96.075127}, yet in the present paper we perform the
derivation by putting bosonic and fermionic dualities on equal footing. In
the rest of the paper, we are going to focus on $p=1$ case.

\section{Bosonic Quantum Hall State}\label{BosonicHallState}

\subsection{Dual descriptions of the bosonic quantum Hall state}

In this section we present the main result, discussing various description of our bosonic theory near filling factor $\nu=1$.
A potential way to realize the bosonic
quantum Hall states is by using the rapidly rotating Bose-Einstein
condensate (BEC) \cite{RevModPhys.81.647}. It is anticipated that some
quantum Hall states will emerge as the Abrikosov lattice melts \cite{PhysRevLett.87.120405}.  
We
will use the flux attachment picture \cite{PhysRevLett.63.199} to
guide our intuition.
   
Particularly interesting is the limiting compressible state at $\nu
=1$. In analogy with the fermionic Halperin-Lee-Read (HLR) $\nu = \frac{1}{2}$ state, this
state has been expected to be a Fermi liquid of the composite
fermions~\cite{PASQUIER1998719, PhysRevB.58.16262}.  On the other
hand, numerical studies on the rotating Bose-Einstein condensate suggested
a gapped ground state \cite{PhysRevLett.87.120405, PhysRevLett.91.030402}, perhaps a Pfaffian
state.
The theory considered here differs from the one describing the rotating
BEC by relativistic invariance and the gauge interactions between
the bosons.  To simplify further discussion we will assume no pairing instability of the composite
fermions, or if there is such an instability we are at a temperature above
the critical temperature.

The minimal ingredient to model the bosonic quantum Hall physics is
the Wilson-Fisher boson defined by~\eqref{WF}. Making use of the
duality web \cite{SEIBERG2016395,PhysRevX.6.031043,PhysRevX.7.041016},
the theory on the particle side $\phi$ is dual to a fermionic theory
\begin{subequations}\label{dualfermion12}
\begin{align}
  I_{\rm m}[\phi, A] & + I_{\rm M}[A;e]\label{dualfermion1l}\\
& \updownarrow\notag\\
I_{\rm m}[\psi, a]-\frac{1}{2}I_{\rm CS}[a]-I_{\rm BF}& [a, A]-I_{\rm CS}[A]+I_{\rm M}[A;e],\label{dualfermion1r}
\end{align}
while on the vortex side $\tilde{\phi}$ one has the duality
\begin{align}
I_{\rm m}[\tilde{\phi},\tilde a]+I_{\rm BF}&[\tilde a,A]+I_{\rm M}[A;e]\label{dualfermion2l}\\
 & \updownarrow\notag \\
I_{\rm m}[\chi, c]+\frac{1}{2}I_{\rm CS}[c]-I_{\rm BF}& [c, A]+I_{\rm CS}[A]+I_{\rm M}[A;e]\label{dualfermion2r}.
\end{align}
\end{subequations}

We define the physical time reversal as the symmetry under which
the Wilson-Fisher boson \eqref{dualfermion1l} is invariant. The naive
time reversal and charge conjugation on the fermionic sides of the
duality will be denoted as $\mathsf T_{\!\rm f}$ and $\mathsf C_{\rm
  f}$.  They act on the gauge fields as 
\begin{align}
  \mathsf T_{\!\rm f}: & ~ \alpha_0 \to \alpha_0,
  ~ \alpha_i \to -\alpha_i ~~ (\alpha=a,c,A), \\
  \mathsf C_{\rm f}: & ~ a_\mu \to -a_\mu,  ~ c_\mu \to -c_\mu,
  ~ A_\mu \to A_\mu, 
\end{align}
and on the fermionic fields $\psi$ and $\chi$ in such a way that the
fermionic kinetic terms are invariant.

There are two apparent puzzles: (i) The
Lagrangians~\eqref{dualfermion1r} and \eqref{dualfermion2r} map to
each other under $\mathsf C_{\rm f}\mathsf T_{\!\rm f}$.  On the other
hand, the mapping between \eqref{dualfermion1l}
and~\eqref{dualfermion2l} is nonlocal.  (ii) Only one of the four
equivalent theories written in Eqs.~\eqref{dualfermion12} [the (2+1)-dimensional Wilson-Fisher boson], is manifestly time reversal
invariant, whereas the other theories seem not to be, at least
classically. These puzzles are explained in
Ref.~\cite{SEIBERG2016395}: the invariance under $\mathsf T$,
manifesting as a classical symmetry of the Wilson-Fisher
boson~\eqref{dualfermion1l}, emerges on the fermionic sides as a
quantum symmetry, which maps a theory to its dual. It has also been
pointed out in Ref.~\cite{PhysRevX.7.041016} that particle-vortex
dualities, under bosonization or fermionization, become local symmetry
operations.

We will see in the following that the naive time-reversal operation on
the fermion side of the duality is analogous to the particle-vortex
duality in the Dirac composite fermion
theory~\cite{PhysRevX.5.031027}. On top of that, after introducing
$I_{\rm M}$ there exists a value of $e^2$ at which the fermionic theories
are $\mathsf T_{\!\rm f}$ invariant.

The four theories defined in Eqs.~(\ref{dualfermion12}) are
equivalent, so a given state can be described in all four theories.  A
state with a certain filling factor $\nu_\phi$ in the original theory
of $\phi$ maps to states with different filling factors in the other
three theories.  We can look at the duality mappings given by these
theories by varying Lagrangians with respect to $A$, $a$, $\tilde a$,
and $c$:
\begin{align}
\label{dualitymapping3} J^{\mu}_{\mathrm{phys}} & = J^{\mu}_{\phi} = -\frac{1}{2\pi}\epsilon^{\mu\nu\lambda}\p_{\nu}a_{\lambda}-\frac{1}{2\pi}\epsilon^{\mu\nu\lambda}\p_{\nu}A_{\lambda}\\
\label{dualitymapping4} =& \frac{1}{2\pi}\epsilon^{\mu\nu\lambda}\p_{\nu}\tilde{a}_{\lambda} = \frac{1}{2\pi}\epsilon^{\mu\nu\lambda}\p_{\nu}A_{\lambda}-\frac{1}{2\pi}\epsilon^{\mu\nu\lambda}\p_{\nu}c_{\lambda},\\
\label{dualitymapping5} J^{\mu}_{\psi} & = \frac{1}{4\pi}\epsilon^{\mu\nu\lambda}\p_{\nu}a_{\lambda}+\frac{1}{2\pi}\epsilon^{\mu\nu\lambda}\p_{\nu}A_{\lambda},\\
 \label{dualitymapping6}J^{\mu}_{\tilde{\phi}} &= -\frac{1}{2\pi}\epsilon^{\mu\nu\lambda}\p_{\nu}A_{\lambda},\\
 \label{dualitymapping7}J^{\mu}_{\chi} & = -\frac{1}{4\pi}\epsilon^{\mu\nu\lambda}\p_{\nu}c_{\lambda}+\frac{1}{2\pi}\epsilon^{\mu\nu\lambda}\p_{\nu}A_{\lambda}.
\end{align}
First, we look at the zeroth components of~\eqref{dualitymapping3} and~\eqref{dualitymapping5}:
\begin{align}
\label{dualitymappingrho3}& \rho_{\phi} = \frac{b}{2\pi}+\frac{B}{2\pi}\Rightarrow \nu_{\phi} = \frac{\lan b\ran}{\lan B\ran}+1,\\
\label{dualitymappingrho4}& \rho_{\psi} = -\frac{b}{4\pi}-\frac{B}{2\pi}.
\end{align}
Therefore, if the original boson has $\nu_{\phi} = 1$, then $\lan b\ran
=0$. The $\psi$ fermions have a finite density $\rho_{\psi} = -\lan
B\ran/(2\pi)$ and live in an average zero magnetic field and
therefore can form a Fermi liquid.

At more general filling fractions, we find 
\begin{align}
\label{fillingmapping2}(\nu_{\phi}-1)\Big(\nu_{\psi}+\frac{1}{2}\Big) = -1.
\end{align}
In particular, if $\psi$ forms an integer quantum Hall state with
$\nu_\psi = n+\frac{1}{2}$, then
\begin{equation}
\label{nuPhi}
 \nu_{\phi} = 1-\frac{2}{2\nu_{\psi}+1} = \frac{n}{n+1},
\end{equation}
which are the Jain sequences for bosons
\cite{PhysRevLett.91.030402}. Eq.~\eqref{nuPhi} is reminiscent of the conventional flux attachment since duality is in essence the relativistic counterpart. Note that if we choose $\nu_\psi =
-(n+\frac12)$, then $\nu_\phi = \frac{n+2}{n+1} =
2-\frac n{n+1}$.  The transition from filling factor $\nu$ to $2-\nu$
is the ``particle-hole'' transformation for bosons, considered in
Ref.~\cite{PhysRevB.94.245107}. Here we see that a symmetry operation
corresponds to the time reversal of composite fermions, which flips
the sign of $\nu_\psi$.  However, since such time reversal is not the
symmetry of the fermionic theory, the physics of the bosonic states
with filling factors $\nu$ and $2-\nu$ are not equivalent.

Straightforwardly, one can make use of Eq.~\eqref{dualitymapping6} to
eliminate $A$ in Eqs.~\eqref{dualitymapping4} and
\eqref{dualitymapping7} and derive the mapping between the filling
fractions in the $\tilde{\phi}$ and the $\chi$ theories. It turns out
the relation is the same as Eq.~\eqref{fillingmapping2},
\begin{align}\label{nuphitilde-nuchi}
 (\nu_{\tilde{\phi}}-1)\Big(\nu_{\chi}+\frac{1}{2}\Big) = -1.
\end{align}

Since the theories involving $\phi$ and $\tilde\phi$ are particle-vortex duals of each other, the connection between the filling factors in the two theories is given by the standard relation
\begin{equation}
\nu_\phi  \nu_{\tilde\phi} = -1,
\end{equation}
and from Eqs.~\eqref{fillingmapping2} and \eqref{nuphitilde-nuchi}, we obtain 
\begin{align}
\nu_{\psi}\nu_{\chi} &= -\frac{1}{4}.
\end{align}

In particular, when the original boson is in the Jain state with
$\nu_\phi=\frac n{n+1}$, the $\chi$ fermion has filling factor
$\nu_\chi=-\frac1{2(2n+1)}$, which is a fractional quantum Hall state
(a Jain state).  The $\nu_\phi=1$ state corresponds to $\nu_\chi=0$,
i.e., the half-filled Landau level of $\chi$. We list some examples in
Table~\ref{table1}.

\begin{table}[h]
\begin{ruledtabular}
\centering
\caption{\label{table1}Some examples of filling fractions under duality mapping.}
\begin{tabular}{cccc}
Field & $\nu$ & $\nu(n=\infty)$ & $\nu(n=0)$\\
$\phi$ & $\frac{n}{n+1}$ & 1 & 0 \\
$\tilde{\phi}$ & $-\frac{n+1}{n}$ & $-1$ & $\infty$\\
$\psi$ & $n+\frac{1}{2}$ & $\infty$ & $\frac{1}{2}$\\
$\chi$ & $\frac{-1}{2(2n+1)}$ & $0$ & $-\frac{1}{2}$\\
\end{tabular}
\end{ruledtabular}
\end{table}

\subsection{Fermionic representations of bosonic observables}

In Sec.~\ref{selfdualtransport} we reviewed relations between
transport coefficients, Eq.~\eqref{result1}, \eqref{result2},
and~\eqref{result3}, in theories that map to each other under
particle-vortex duality.  We can also derive the connection between
the transport coefficients between the bosonic and fermionic sides of
each duality \eqref{dualfermion12}.  We write down the spatial
components of Eqs.~\eqref{dualitymapping3} and \eqref{dualitymapping5}
and introduce the relevant transport coefficients,
\begin{align}
& \mb J_{\phi} = \sigma_{\phi}\mb E+\alpha_{\phi}\bm{\nabla} T = \frac{1}{2\pi}\epsilon\,\mb e+\frac{1}{2\pi}\epsilon\mb E,\\
& \mb J_{\psi} = \sigma_{\psi}\mb e+\alpha_{\psi} \bm{\nabla} T = -\frac{1}{4\pi}\epsilon\,\mb e-\frac{1}{2\pi}\epsilon\mb E.
\end{align} 
In the same manner as in Ref.~\cite{PhysRevB.95.045118}, the consistency between the two equations requires
\begin{subequations}\label{fbresult124}
\begin{align}
\label{fbresult1}& \epsilon\Big(\sigma_{\phi}-\frac{\epsilon}{2\pi}\Big)\epsilon^{-1}\Big(\sigma_{\psi}+\frac{\epsilon}{4\pi}\Big) = \frac{1}{(2\pi)^2},\\
\label{fbresult2}& \alpha_{\phi} = -\frac{\epsilon}{2\pi}\Big(\sigma_{\psi}+\frac{\epsilon}{4\pi}\Big)^{-1}\alpha_{\psi}.
\end{align}
In addition to electrical and thermoelectric responses, we further look at thermal transport.  As the heat current should have the same form in either of the dual theories,
$\mb q = -\bar{\kappa}_{\phi}\nabla T-\alpha_{\phi}T\mb E = -\bar{\kappa}_{\psi}\nabla T-\alpha_{\psi} T\mb e$,
we can then obtain 
\begin{align}
   \label{fbresult4}\bar{\kappa}_{\phi} = \bar{\kappa}_{\psi}-\alpha_{\psi}\Big(\sigma_{\psi}+\frac{1}{4\pi}\Big)^{-1}\alpha_{\psi}T.
\end{align}
\end{subequations}
It is straightforward to show these relationships hold with the replacement $\phi\to\tilde{\phi}$ and $\psi\to \chi$.

One can also verify that if the transport coefficients in the two bosonic ($\phi$ and $\tilde\phi$) theories are related by 
Eqs.~\eqref{result123} with $p=1$, then Eqs.~\eqref{fbresult124} (and similar equations with the replacement $\phi\to\tilde{\phi}$ and $\psi\to \chi$)  imply that the transport coefficients in the two fermionic theories satisfy the duality relations with $p=2$.  This is consistent with 
 $\psi$ and $\chi$ being related by particle-vortex duality. 

\subsection{Transport in self-dual boson:A fermionic view}

We now look at the self-dual $\nu_{\phi} =1$ state and show that the constraints that follow from duality can be understood using the fermionic picture, in which $\psi$ forms a Fermi surface.  At self-duality $(\sigma_\phi^{xx})^2 +(\sigma_\phi^{xy})^2=\frac1{(2\pi)^2}$, and one can parametrize $\sigma_{\phi}^{xx} =\frac{1}{2\pi}\cos\theta_H $ and $\sigma_{\phi}^{xy} = \frac{1}{2\pi}\sin\theta_H$, where $\theta_H$ is the Hall angle. Using Eq.~\eqref{fbresult1}, we have 
\begin{align}
\label{fbresult5}& \sigma_{\psi}^{xx} = \frac{1}{4\pi} \tan
\Big(\frac{\theta_H}{2}+\frac{\pi}{4}\Big)
,\quad   \sigma_{\psi}^{xy} = 0.
\end{align}
In reverse, if the Hall conductivity of the composite fermion $\sigma_\psi^{xy}$ vanishes, then the bosonic conductivity tensor satisfies the self-duality constraint.

Note that the average magnetic field acting on the composite fermion is zero, and the vanishing of $\sigma_\psi^{xy}$ takes place in the simple Drude model of transport.  It is instructive to derive all three self-duality constraints for bosonic transport from the Drude model of the composite fermion.  This model gives only diagonal transport tensors
\begin{align}
\label{Drudeassump}\sigma_{\psi}^{ij} = \sigma_{\psi}\delta^{ij},\ \alpha_{\psi}^{ij} = \alpha_{\psi}\delta^{ij},\ \kappa^{ij}_{\psi} = \kappa_{\psi}\delta^{ij}.
\end{align}
Plugging $\sigma^{ij}_{\psi}$ into Eq.~\eqref{fbresult1}, we find
\begin{align}
& \sigma_{\phi}^{xx} = \frac{1}{4\pi^2}\frac{\sigma_{\psi}^{xx}}{(\sigma_{\psi}^{xx})^2+\frac1{(4\pi)^2}},\notag\\
& \sigma_{\phi}^{xy} = \frac{1}{2\pi}-\frac{1}{16\pi^3}\frac{1}{(\sigma_{\psi}^{xx})^2+\frac1{(4\pi)^2}} \notag\\
\label{fbresult7}\Rightarrow & (\sigma_{\phi}^{xx})^2+(\sigma_{\phi}^{xy})^2 = \frac{1}{(2\pi)^2}.
\end{align}
Given the above, we can parametrize $\sigma_{\phi}$ using the Hall angle. Then by the same token, using Eqs.~\eqref{fbresult2} and~\eqref{fbresult5}, we find 
\begin{align}
\label{fbresult8}\frac{\alpha^{xy}_{\phi}}{\alpha^{xx}_{\phi}} = 4\pi\sigma_{\psi}^{xx} = \tan\Big(\frac{\pi}{4}+\frac{\theta_H}{2}\Big).
\end{align} 
Finally, we look at the difference $\bar{\kappa}^{xy}_{\phi}-\bar{\kappa}^{yx}_{\phi}$. Using Eqs.~\eqref{fbresult4} and~\eqref{fbresult5}, it is
\begin{align}
\bar{\kappa}_{\phi}^{xy}-\bar{\kappa}_{\phi}^{yx} = \frac{8\pi T\alpha_{\psi}^2}{(16\pi^2\sigma_{\psi}^2+1)}= 8\pi T\alpha_{\psi}^2\cos^2\Big(\frac{\pi}{4}+\frac{\theta}{2}\Big).\notag
\end{align}
To find $\alpha^2_{\psi}$ in terms of $\alpha_{\phi}$, we use~\eqref{fbresult2} and~\eqref{fbresult8} to get 
\begin{align}
\alpha^2_{\psi} =& \frac{1}{4}\sec^2\Big(\frac{\pi}{4}+\frac{\theta}{2}\Big)\left[(\alpha_{\phi}^{xx})^2+(\alpha_{\phi}^{xy})^2\right],\notag
\end{align}
and as a consequence, 
\begin{align}
\label{fbresult9}\bar{\kappa}^{xy}_{\phi}-\bar{\kappa}^{yx}_{\phi} = 2\pi T \left[(\alpha_{\phi}^{xx})^2+(\alpha_{\phi}^{xy})^2\right].
\end{align}
To summarize, the exact relationships \eqref{result4},~\eqref{result5}, and~\eqref{result6}, which are the consequences of self-duality in the bosonic theory, can be derived assuming~\eqref{Drudeassump} on the fermion side, which would appear naturally in the simplest Drude model of transport.


The above ``derivation'' of the self-duality constraints on bosonic transport gives rise to a puzzle. 
If we recall the derivations in the self-dual QEDs, these identities hold under two assumptions: (i) $e^2 = 4\pi p$, and (ii) the matter field is tuned at self-dual filling $\nu = 1/p$.  On the other hand, the derivation through the fermionic Drude model does not seem to requires one to tune $e^2 = 4\pi$.  In the literature, a similar argument was used \cite{PhysRevB.95.045118} to derive the semicircle law (\ref{result4}) in the absence of any self-duality.
 In the following section, we explain why the fine tuning of $e^2$ is required for the derivation to work.

\subsection{Self-duality and manifest $\mathsf T_{\!\rm f}$ symmetry}\label{tsymmetry}

In the previous section, we showed that all self-dual properties of boson field $\phi$ can be understood in terms of those of $\psi$ at the level of Drude approximation. This argument does not make use of the value of $e$ and thereby is independent of self-dual structure, implying that the exact relationships between transport coefficients that we have derived using self-duality are valid even when the coupling constant is away from the self-dual point. 
 In this section we show that the naive argument is not correct and, indeed, the fine tuning of the coupling constant is required.

\begin{figure}
\includegraphics[width = 1.0\columnwidth]{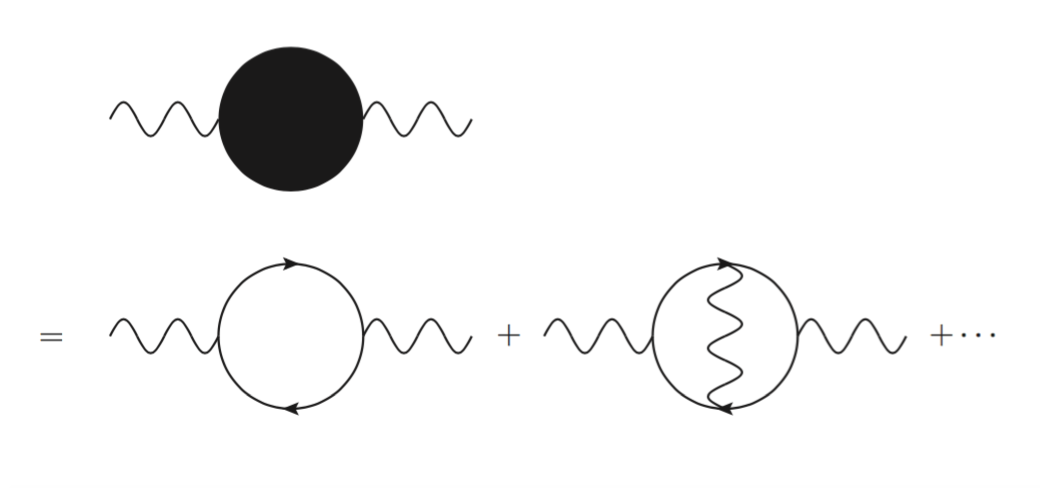}
\caption{\label{fig1} In the present work, transport coefficients are defined as the 1PI diagram in the first line. In the second line, the leading diagram on the left is given by the RPA, which can be isotropic. As higher-order diagrams such as the right one in the second line are included, $\lan aa\ran$ will contribute to parity-violating off-diagonal terms at general $e^2$ and invalidate the assumption of Drude values.}
\end{figure}

We note that those transport properties hold as long as $(\sigma_{\psi}, \alpha_{\psi}, \bar{\kappa}_{\psi}, \kappa_{\psi})$ have no off-diagonal and parity-breaking components and are isotropic, being proportional to $\delta^{ij}.$  We have also argued that, since the average magnetic field is zero, Drude approximation indeed gives  zero off-diagonal transport coefficients.  However, the approximation neglects gauge field fluctuations.  In a theory where the dynamics of the gauge field violates parity and time-reversal symmetries, a contribution of the type depicted in the two-loop self-energy diagram in Fig.~\ref{fig1} necessarily introduces nonzero Hall transport. 

It remains for us to explain why at a particular value of the gauge coupling, namely, the self-dual value $e^2=4\pi$, gauge field fluctuations do not lead to nondiagonal transport coefficients.

To do that, we integrate over $A$ in Eq.~\eqref{dualfermion1r} and obtain an effective action with a nonlocal term for $a$ (a pseudo quantum electrodynamics (PQED) in the terminology of Ref.~\cite{Marino:1992xi}
\begin{align}
\label{Ieff} & I_{\mathrm{eff}} = I_{\rm m}[\psi, a]+\frac{e^2-\tilde e^2}{2(e^2+\tilde e^2)}I_{\rm CS}[a]+I_{\rm P}[a; g]\\
& \frac{1}{g^2} = \frac{1}{e^2}\frac{1}{[1+(4\pi/e^2)^2]} = \frac{1}{e^2+\tilde e^2}.
\end{align}
We see that at $e^2 = 4\pi$, the integration procedure completely cancels out $-\frac{1}{8\pi}a\, d a$ in the original Lagrangian. This provides an explanation for the puzzle: only at this value of $e^2$ would the off-diagonal transport coefficient in the fermionic theory remain zero once one goes beyond the mean-field approximation and integrates over gauge field fluctuations. 

Note that the coincidence between self-dual and time-reversal restoration is also found and discussed in the wire-construction version of the model \cite{PhysRevX.7.041016}.

\subsection{Electromagnetic duality revisited}
In Sec.~\ref{tsymmetry} we saw that, as $e^2 = 4\pi$, the fermionic theories are also tuned to their self-dual point. It is instructive to examine this fact from the point of view of electromagnetic duality in the bulk. In this section we show that at $e^2=4\pi$, the gauge field seen by $\psi$ is the electromagnetic dual of one seen by $\chi$. To avoid confusion, we introduce an index to distinguish the gauge field that couples to the matter field in each of the four theories, so
\begin{equation}
\begin{split}
  &\phi ~ \textrm{couples to}  ~ A, \quad  \psi ~\textrm{couples to} ~ a_{\psi},\\
  & \tilde{\phi} ~\textrm{couples to} ~ a_{\tilde{\phi}},\quad  \chi ~\textrm{couples to} ~ a_{\chi}. 
\end{split}
\end{equation}

For a gauge field whose action contains a (3+1)D Maxwell term $I_{\rm M}$, one can impose certain orbifold conditions~\cite{PhysRevB.96.075127}. For example, in the theory of $\phi$, one can require 
\begin{equation}
\begin{split}
   A_{\mu}(z) &= A_{\mu}(-z), \quad \mu=t,x,y ,\\
   A_{z}(z) &= -A_z(-z)
\end{split}
\end{equation}
Such orbifold conditions can be imposed without changing the theory because the parts of $A$ with opposite parities (odd part of $A_\mu$ and even part of $A_z$) decouple~\cite{PhysRevB.96.075127}. 

We first relate fields $A$ and $a_{\psi}$. Using the boundary condition for the bulk Maxwell equation and the relations between $J^{\mu}_{\phi}$, $a_{\psi}$, and $A$, Eqs.~\eqref{dualitymapping3} and \eqref{dualitymapping5}, we have 
\begin{subequations}\label{EbB}
\begin{align}
& \frac{2\pi}{e^2}\Delta E_z 
=\frac{4\pi}{e^2}E_z( 0^+) = b_{\psi}+B_z\\
& \frac{2\pi}{e^2}\Delta\mb B_{\parallel} = \frac{4\pi}{e^2}\mb B_{\parallel}( 0^+) = -\mb e_{\psi}-\mb E_{\parallel}.
\end{align}
\end{subequations}
In the formulas that follow, quantities that are discontinuous across the plane $z=0$ will be assumed to be evaluated at $z=+0$ unless we explicitly specify otherwise.
Next, we perform an electromagnetic duality transformation in the bulk:
\begin{align}
  & \mb E = -\mathrm{sgn}(z)\tilde{\mb B},\\
  & \mb B = \mathrm{sgn}(z)\tilde{\mb E}.
\end{align}
Note that we have chosen opposite sign conventions in the definition of the dual gauge fields $\mb E$ and $\mb B$ on the two sides of the brane. 
Equations~(\ref{EbB}) now can be rewritten as
\begin{align}
 -\frac{4\pi}{e^2}\tilde B_z & = b_{\psi}+\tilde E_z,\\
 \frac{4\pi}{e^2}\tilde{\mb E}_{\parallel} &= -\mb e_{\psi}+\tilde{\mb B}_{\parallel}.
\end{align}
As suggested by this relation, in the dual $\psi$ theory, we extend gauge fields $a_{\psi}$ into the bulk by 
\begin{align}\label{relationetoE1}
\begin{pmatrix}
\mb e_{\psi} \\ \mb b_{\psi} 
\end{pmatrix}= \frac{1}{e}\begin{pmatrix} -\tilde e & e\\ -e&  -\tilde e\end{pmatrix}\begin{pmatrix} \tilde {\mb E}\\ \tilde{ \mb B}\end{pmatrix},\ \tilde e = \frac{4\pi}{e}.
\end{align}
The relation between $A$ and $a_{\tilde{\phi}}$ was shown previously in Ref. \onlinecite{PhysRevB.96.075127}. Because of particle vortex duality, 
\begin{align}
\label{relation7}&\rho_{\tilde{\phi}} = \frac{1}{2\pi}B_z = \frac{1}{2\pi}\tilde{E}_z(0^+),\\
\label{relation8}& \mb J_{\tilde{\phi}} = \frac{1}{2\pi}\epsilon\mb E_{\parallel} = -\frac{1}{2\pi}\epsilon \tilde{\mb B}_{\parallel}(0^+).
\end{align}
$a_{\tilde{\phi}}$ is identified on the brane and extended to the bulk with $A$ (or $\tilde A$) via 
\begin{align}
a_{\tilde{\phi}} = \frac{4\pi}{e^2}\tilde A.
\end{align}
Finally, we relate $a_{\chi}$ and $A$ using~\eqref{dualitymapping4},~\eqref{dualitymapping6},~\eqref{dualitymapping7}, \eqref{relation7}, and~\eqref{relation8}. After eliminating $J_{\tilde{\phi}}^{\mu}$, in terms of $\tilde A$, 
\begin{align}
& \tilde{E}_z = b_{\chi}+\frac{4\pi}{e^2}\tilde{B}_z,\\
& \tilde{\mb B}_{\parallel} = -\mb e_{\chi}-\frac{4\pi}{e^2}\tilde{\mb E}_{\parallel}.
\end{align}
Again, as suggested by this relation, in $\chi$ theory we extend $a_{\chi}$ into the bulk by 
\begin{align}\label{relationetoE2}
\begin{pmatrix}
\mb e_{\chi} \\ \mb b_{\chi} 
\end{pmatrix}= \frac{1}{e}\begin{pmatrix} -\tilde e & -e\\ e&  -\tilde e\end{pmatrix}\begin{pmatrix} \tilde {\mb E}\\ \tilde{ \mb B}\end{pmatrix}.
\end{align}
Now we can relate $a_{\psi}$ and $a_{\chi}$ by these relations:
\begin{align}
\label{relation10}\begin{pmatrix} \mb e_{\chi} \\ \mb b_{\chi}\end{pmatrix} = \frac{1}{e^2+\tilde e^2}\begin{pmatrix} \tilde e^2-e^2 & 8\pi \\ -8\pi & \tilde e^2-e^2\end{pmatrix}\begin{pmatrix} \mb e_{\psi} \\ \mb b_{\psi}\end{pmatrix}.
\end{align}
Therefore, at $e^2 = 4\pi$, 
\begin{align}
& \mb e_{\chi} = \mb b_{\psi},\\
& \mb b_{\chi} = -\mb e_{\psi}.
\end{align}
To summarize, at any $e^2$, $a_{\phi}$ and $A$ are related by electromagnetic duality and $a_{\psi}$ and $a_{\chi}$ can be related by~\eqref{relation10}. However, at $e^2 = 4\pi$, where $\phi$ and $\tilde{\phi}$ becomes self-dual, $a_{\chi}$ and $a_{\psi}$ also become the electromagnetic duals of each other.\\
Continuing this procedure, we may look at the effective action led by this integration process. Using duality mapping~\eqref{dualitymapping5}, 
\begin{align}
& \rho_{\psi}  =-\frac{b_{\psi z}}{4\pi}-\frac{\tilde E_z}{2\pi}, \\
& \mb J_{\psi} = -\frac{1}{4\pi}\mb e_{\psi\parallel}\times\hat{\mb z}+\frac{1}{2\pi}\tilde{\mb B}_{\parallel}\times\hat{\mb z}.
\end{align}
Inverting~\eqref{relationetoE1}, the above equations become 
\begin{align}
& e_{\psi z} = \frac{e^2+\tilde e^2}{2}\Big(\rho_{\psi}-\frac{1}{4\pi^2}\frac{(e^2-\tilde e^2)\pi}{e^2+\tilde e^2}b_{\psi z}\Big),\\
& \mb b_{\psi\parallel} = \frac{e^2+\tilde e^2}{2}\Big(\mb J_{\psi}\times\hat{\mb z}+\frac{1}{4\pi^2}\frac{(e^2-\tilde e^2)\pi}{e^2+\tilde e^2}\mb e_{\psi\parallel}\Big).
\end{align}
Similarly, inverting~\eqref{relationetoE2} gives us 
\begin{align}
& e_{\chi z} = \frac{e^2+\tilde e^2}{2}\Big(\rho_{\chi}+\frac{1}{4\pi^2}\frac{(e^2-\tilde e^2)\pi}{e^2+\tilde e^2}b_{\chi z}\Big),\\
&  \mb b_{\chi\parallel} = \frac{e^2+\tilde e^2}{2}\Big(\mb J_{\chi}\times\hat{\mb z}-\frac{1}{4\pi^2}\frac{(e^2-\tilde e^2)\pi}{e^2+\tilde e^2}\mb e_{\chi\parallel}\Big).
\end{align}
Referring to the general theory of axion electrodynamics \cite{PhysRevLett.58.1799}, these boundary conditions are given by the actions 
\begin{align}
&I_{\rm m}[\psi, a_{\psi}]+I_{\rm M}[a_{\psi}, g]+\frac{1}{4\pi^2}\int d^4x\, \theta(z) \mb e_{\psi}\cdot\mb b_{\psi}.\\
& I_{\rm m}[\chi, a_{\chi}]+I_{\rm M}[a_{\chi},g]-\frac{1}{4\pi^2}\int d^4x\, \theta(z) \mb e_{\chi}\cdot\mb b_{\chi}\\
& g^2 = e^2+\tilde e^2, \theta(z) = \frac{\mathrm{sgn}(z)}{2}\Big(\frac{e^2-\tilde e^2}{e^2+\tilde e^2}\Big)\pi.
\end{align}
The choice of the $\theta$ term accommodates the orbifold boundary conditions. From this perspective, we learn that in the presence of the bulk fluctuation, bosonization dualities have the following structure:
\begin{align}
& I_{\rm m}[\phi, A]+I_{\rm M}[A;e]\leftrightarrow I_{\rm m}[\psi, a]+I_{\rm M}[a;g]+I_{\theta}[a; \Delta\theta]\notag\\
&\ \ \ \ \ \ \ \ \ \ \ \updownarrow\ \ \ \ \ \ \ \ \ \ \ \ \ \ \ \ \ \ \ \ \ \ \ \ \ \ \ \ \ \ \ \ \ \updownarrow\notag\\
& I_{\rm m}[\tilde{\phi}, \tilde A] + I_{\rm M}[\tilde A, \tilde e]\leftrightarrow I_{\rm m}[\chi, a]+I_{\rm M}][a;g]+I_{\theta}[a; -\Delta\theta]\notag
\end{align}
In terms of the bulk field actions, the manifest $\mathsf T$-symmetric point occurs when we flow $e^2$ to $\tilde e^2$, tuning $\Delta\theta = 0$.

\section{Modular Invariance}

We have defined the mixed-dimensional QED so that the gauge field
propagates on both sides of the (2+1)D brane.  In
Ref.~\cite{PhysRevB.96.075127} we compared this picture with the
alternative picture often used in the literature (for example, in
Ref.~\cite{SEIBERG2016395}), where the brane is the boundary of space
and the bulk fields propagate only on one side of it.  There we see that
the self-dual structure is also evident in the latter model, in which
fermion and boson self-dual couplings are $g_{\mathrm f}^2 = 4\pi$ and
$g_{\mathrm b}^2 = 2\pi$, respectively.

Let us argue that the fermionic self-dual point $g_{\mathrm f}^2 = 4\pi$ is
equivalent to $g_{\mathrm b}^2 = 2\pi$ from this perspective. More
concretely, we consider a bulk electromagnetic action with complex
coupling constant $\tau$ with a boundary matter minimally coupled to
the bulk field. The mixed Chern-Simons term $a\, d A$ and Chern-Simons
term $A\, d A$ are regarded as the results of $\mathsf S$ and $\mathsf
T$. The action of deforming action, via a Gaussian integral, is
rephrased in terms of mapping coupling constants via
\cite{1126-6708-1999-04-021, Witten:2003ya}
\begin{align}
& \mathsf S[\tau] = -\frac{1}{\tau}\\
& \mathsf T[\tau] = \tau+1.
\end{align}
We denote the bulk action as $I_{\mathcal M}[\tau]$ and the boundary
minimally coupled matter action as $I_{\mathrm{b/f}}[\cdots]$. As a
warm-up, we may recall the fermionic self dual theories can be stated
concisely as
\begin{align}
I_{\mathrm f}[\psi]+I_{\mathcal M}[\tau_{\mathrm f}]\leftrightarrow I_{\mathrm f}[\tilde{\psi}]+I_{\mathcal M}[\tilde\tau_{\mathrm f}],
\end{align}
with 
\begin{align}
\tilde\tau_{\mathrm f}= \mathsf S\mathsf T^{-2}\mathsf S\mathsf T^{-1}[\tau_{\mathrm f}] = \frac{\tau_{\mathrm f}-1}{2\tau_{\mathrm f}-1}.
\end{align}
Similarly, the bosonic self-dual theories are stated as 
\begin{align}
I_{\mathrm b}[\phi]+ I_{\mathcal M}[\tau_{\mathrm b}]\leftrightarrow I_{\mathrm b}[\tilde\phi]+I_{\mathcal M}[\tilde\tau_{\mathrm b}],
\end{align}
with 
\begin{align}
\tilde\tau_{\mathrm b} = \mathsf S[\tau_{\mathrm b}] = -\frac{1}{\tau_{\mathrm b}}.
\end{align}
The boson-fermion duality states that 
\begin{align}
I_{\mathrm b}[\phi]+ I_{\mathcal M}[\tau_{\mathrm b}]\leftrightarrow I_{\mathrm f}[\psi]+I_{\mathcal M}[\tau_{\mathrm f}],
\end{align}
where \footnote{It may differ by a $\mathsf T$ operation depending how we choose the real part of $\tau_{\mathrm f}$.} 
\footnote{More precisely, this is the relation between $\phi$ and $\chi$ coupling constants.}
\begin{align}
\label{taumapping}\tau_{\mathrm f} = \frac{1}{1-\tau_{\mathrm b}}, \ \mathsf T\mathsf S[\tau_{\mathrm f}] = \tau_{\mathrm b}. 
\end{align}
At the bosonic self-dual point $\tau_{\mathrm b} = \tilde\tau_{\mathrm b}
= i$, we see fermion theories are also pinned at the self-dual point
$\tau_{\mathrm f} = \tilde\tau_{\mathrm f} = \frac{1}{2}(1+i)$. To
avoid possible confusion, we note that up to a $\mathsf T$ operation
the self-dual coupling of fermions can also be
\begin{align}
\tau_{\mathrm f} = -\frac{1}{2}+\frac{i}{2}.
\end{align}
This picture can provide additional understanding of the fact the parity-violating term in the effective action vanishes at the self-dual
point. The relation between $\tau_{\phi}$ and $\tau_{\psi}$ is given
by
\begin{align}
\tau_{\phi} = -\frac{1}{\tau_{\psi}}-1 = \mathsf T^{-1}\mathsf S[\tau_{\psi}].
\end{align}
Taking $\tau_{\phi} = i\tau_{\phi}''$,
\begin{align}
\tau_{\psi} = -\frac{1}{1+i \tau_{\phi}''} = -\frac{1}{1+(\tau''_{\phi})^2}(1-i \tau_{\phi}'').
\end{align} 
The parity-invariant condition 
\begin{align}
\mathsf{Re}[\tau_{\psi}] = -\frac{1}{2}\Rightarrow \tau_{\phi}'' = 1,\ \mathsf{Im}[\tau_{\psi}] = \frac{1}{2}.
\end{align}
The same argument applies in the reverse, where
\begin{align}
\tau_{\psi} = -\frac{1}{\tau_{\phi}+1} = \mathsf S\mathsf T[\tau_{\phi}].
\end{align}
Taking $\tau_{\psi} = -\frac{1}{2}+i \frac{1}{2}\tau_{\psi}''$,
\begin{align}
\tau_{\phi}  = \frac{1-(\tau_{\psi}'')^2}{1+(\tau_{\psi}'')^2}+\frac{2i }{1+(\tau_{\psi}'')^2}.
\end{align}
The parity-invariant condition 
\begin{align}
\mathsf{Re}[\tau_{\phi}] = 0\Rightarrow \tau_{\psi}'' = 1, \tau_{\phi} = i.
\end{align}
As a simple exercise, we can also redefine the coupling constants such
that $\tau_{\phi}$ and $\tau_{\psi}$ share the same self-dual point as
found in Refs.~\cite{PhysRevX.7.041016, PhysRevB.97.195112}.  Explicitly,
we define $-1/z_{\mathrm f} = 2\tau_{\mathrm f}-1$ and $z_{\mathrm b} =
\tau_{\mathrm b}$. Equation~\eqref{taumapping} becomes
\begin{align}
z_{\mathrm f} = \frac{z_{\mathrm b}-1}{z_{\mathrm b}+1}.
\end{align}
The self-dual point is moved to $z_{\mathrm f} = z_{\mathrm b} = i $.

\section{Conclusion}

To summarize, In the first half of the paper we reviewed the self-dualities led by combining (2+1)-dimensional particle-vortex dualities and 3+1 U(1) gauge fields and conjectured a relation with $\mathsf{S}$ duality. Those dualities also imply mappings between transport properties. In particular, at self-dual points and self-dual states, these mappings become more concrete constraints on the entries of transport tensors.
In the second part, we studied the $\nu = 1$ bosonic quantum Hall state using a description in terms of a
 Dirac composite fermion.  We showed that the self-duality constraints on transport in the bosonic theory can be understood very easily as a consequence of the discrete symmetry of the composite fermion.  We hope that the model studied in this paper will provide additional insights into the composite fermions in quantum Hall states.  

\medskip

\acknowledgments
W.-H.H. thanks O. Motrunich and D. Mross for explaining their insightful work \cite{PhysRevX.7.041016} and stimulating discussion and all the participants at  2018 Aspen Winter Conference on ``Field Theory Dualities and Strongly Correlated Matter'' for insightful comments.  This work is supported, in part, by U.S. Department of Energy Grant No. DE-FG02-13ER41958, Army Research Office Multidisciplinary University Research Initiative Grant No. 63834-PH-MUR, and a Simons Investigator Grant from the Simons Foundation.
\bibliography{citation}{}
\bibliographystyle{apsrev4-1}
\end{document}